\documentclass[twocolumn,prb,aps,amsfonts,amssymb,amsmath,showpacs,superscriptaddress,letterpaper]{revtex4}
\usepackage{graphicx}
\usepackage{hyperref}

\begin{document}

\title{Formation of quantum dots in the potential fluctuations of InGaAs heterostructures probed by scanning gate microscopy}

\author{P. Liu}
 \altaffiliation{Present address: School of Innovation \& Entrepreneurship, Peking University, Beijing, China}
 \affiliation{Universit\'e Grenoble Alpes, Institut NEEL, F-38000 Grenoble, France}
 \affiliation{CNRS, Institut NEEL, F-38042 Grenoble, France}
\author{F. Martins}
 \affiliation{IMCN/NAPS, Universit\'e catholique de Louvain, B-1348 Louvain-la-Neuve, Belgium}
\author{B. Hackens}
 \affiliation{IMCN/NAPS, Universit\'e catholique de Louvain, B-1348 Louvain-la-Neuve, Belgium}
\author{L. Desplanque}
 \affiliation{IEMN, UMR CNRS 8520, UST Lille, F-59652 Villeneuve d'Ascq, France}
\author{X. Wallart}
 \affiliation{IEMN, UMR CNRS 8520, UST Lille, F-59652 Villeneuve d'Ascq, France}
\author{M.G. Pala}
 \affiliation{IMEP-LAHC, Grenoble INP, Minatec, F-38016 Grenoble, France}
\author{S. Huant}
 \affiliation{Universit\'e Grenoble Alpes, Institut NEEL, F-38000 Grenoble, France}
 \affiliation{CNRS, Institut NEEL, F-38042 Grenoble, France}
\author{V. Bayot}
 \affiliation{Universit\'e Grenoble Alpes, Institut NEEL, F-38000 Grenoble, France}
 \affiliation{CNRS, Institut NEEL, F-38042 Grenoble, France}
 \affiliation{IMCN/NAPS, Universit\'e catholique de Louvain, B-1348 Louvain-la-Neuve, Belgium}
\author{H. Sellier}
 \email{hermann.sellier@neel.cnrs.fr }
 \affiliation{Universit\'e Grenoble Alpes, Institut NEEL, F-38000 Grenoble, France}
 \affiliation{CNRS, Institut NEEL, F-38042 Grenoble, France}

\date{\today}

\begin{abstract}

The disordered potential landscape in an InGaAs/InAlAs two-dimensional electron gas patterned into narrow wires is investigated by means of scanning gate microscopy. It is found that scanning a negatively charged tip above particular sites of the wires produces conductance oscillations that are periodic in the tip voltage. These oscillations take the shape of concentric circles whose number and diameter increase for more negative tip voltages until full depletion occurs in the probed region. These observations cannot be explained by charging events in material traps, but are consistent with Coulomb blockade in quantum dots forming when the potential fluctuations are raised locally at the Fermi level by the gating action of the tip. This interpretation is supported by simple electrostatic simulations in the case of a disorder potential induced by ionized dopants. This work represents a local investigation of the mechanisms responsible for the disorder-induced metal-to-insulator transition observed in macroscopic two-dimensional electron systems at low enough density.

\end{abstract}



\maketitle

\section{Introduction}

Two-dimensional electron gases~\cite{ando-82-rmp} (2DEGs) buried inside semiconductor heterostructures~\cite{harris-89-rpp} show ballistic transport over micrometers at low temperature. Their very long electron mean free path results from the combination of a high growth quality by molecular beam epitaxy and a remote doping technique that drastically reduces scattering by impurities.~\cite{stormer-83-ss} In such heterostructures, conduction electrons are confined at the interface between two different band gap materials and spatially separated from the dopants, which are placed a few tens of nanometers above the heterojunction. However, the random distribution of the ionized dopants produces long-range potential fluctuations in the 2DEG that strongly affect electron transport at low temperature.~\cite{rorison-88-sst}

Below a critical electron density, this disorder potential breaks the 2DEG into several electron puddles,~\cite{ilani-01-sci} and conduction is described by a percolation process in a two-dimensional network with thermally activated hopping.~\cite{shi-02-prl} This 2D metal-to-insulator transition (MIT) has been extensively studied by transport experiments in macroscopic samples using large planar gates to control the overall electron density.~\cite{dassarma-05-prl} Investigations of the MIT in small samples revealed that long-range and short-range disorder potentials produce different behaviors. In particular, insulating samples with short gate length show a metallic behavior at very low temperature~\cite{baenninger-08-prl} that may result from resonant tunneling between conducting domains.~\cite{neilson-10-prb} In samples with even shorter gate length, strong conductance fluctuations are observed versus gate voltage~\cite{washburn-88-prb} due to sample specific disorder configurations.~\cite{davies-89-prb}

These potential fluctuations also explain the tremendous difficulty to fabricate ballistic one-dimensional wires.~\cite{nixon-90-prb,laughton-91-prb} The presence of potential barriers along the wire results in the formation of localized states with Coulomb blockade, especially in long wires of several microns in length,~\cite{scott-thomas-89-prl,meirav-89-prb,field-90-prb,staring-92-prb} but also in submicrometer-length wires.~\cite{nicholls-93-prb,poirier-99-prb} In quantum point contacts (QPCs), the presence of potential fluctuations in the constriction~\cite{nixon-91-prb} is often invoked to explain resonances in the quantized conductance plateaus.~\cite{faist-90-prb} Alternatively, QPCs could be used to probe locally the disorder potential, since only a small region between the split-gates dominates the transport.~\cite{larkin-94-prb} Finally, in mesoscopic devices of intermediate dimension at very low temperature, quantum interferences of electron waves spreading coherently in the disordered potential landscape give rise to universal conductance fluctuations~\cite{lee-85-prl,vanhouten-86-apl} (UCFs).

Imaging the disorder potential of a surface 2DEG can be achieved by scanning tunneling microscopy,~\cite{morgenstern-02-prl,wiebe-03-prb} but the case of 2DEGs buried tens of nanometers below the surface requires specific local probe techniques. Most of the studies have been done in the quantum Hall regime at high magnetic field, using techniques based on subsurface charge accumulation,~\cite{tessmer-98-nat,finkelstein-00-sci,steele-05-prl} single electron transistors,~\cite{yoo-97-sci,zhitenev-00-nat,ilani-04-nat} and scanning gate microscopy (SGM).~\cite{woodside-01-prb,baumgartner-07-prb,paradiso-11-prb,martins-13-njp} Surprisingly, very few studies have been done at zero magnetic field. Scanning capacitance microscopy is the only technique that succeeded in imaging directly the disorder potential at zero field and revealed fluctuations on a length scale much larger than expected from the distance between dopants and the 2DEG.~\cite{chakraborty-04-prb} However, SGM can also provide indirect information on the disorder potential inside or close to a nanoscale device by imaging, for example, the complex branched electron flow spreading in a 2DEG out of a QPC,~\cite{topinka-01-nat,jura-07-natp,kozikov-13-njp,brun-14-ncom} the UCF pattern in a small constriction etched in a 2DEG,~\cite{aoki-05-apl,dacunha-06-apl} the irregular fringe pattern in a quantum ring,~\cite{pala-08-prb,pala-09-nt,chwiej-13-prb} or the presence of charge traps in the 2DEG heterostructure.~\cite{pioda-07-prb,gildemeister-07-jap}

\begin{figure}[b]
\begin{center}
\includegraphics[width=7cm,clip,trim=0 0 0 0]{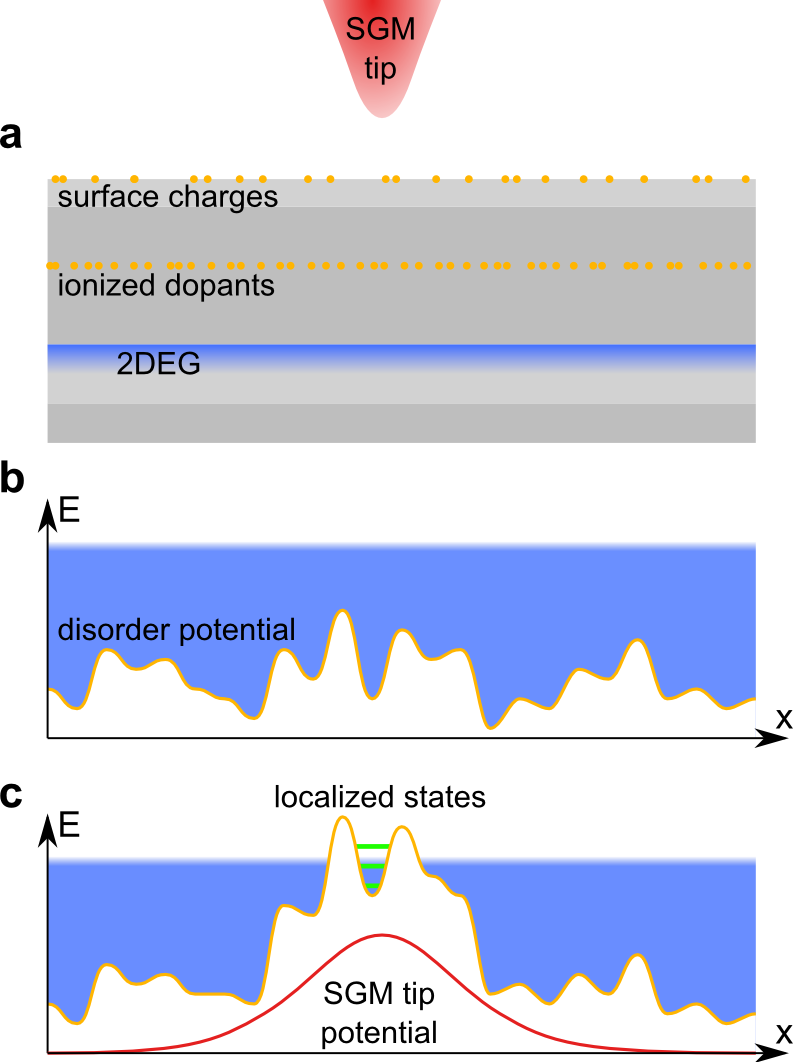}
\caption{(a) Cross-section of the semiconductor heterostructure hosting a high-mobility 2DEG at the heterojunction. (b) Schematics of the energy potential landscape in the 2DEG resulting from the random distribution of ionized dopants and surface charges. (c) Potential landscape in the presence of a negatively polarized SGM tip that raises the potential fluctuations around the Fermi level and creates a quantum dot.} \label{fig1}
\end{center}
\end{figure}

In this paper, we use SGM~\cite{sellier-11-sst,ferry-11-sst} to probe the disorder potential in a low-density InGaAs/InAlAs 2DEG, patterned by etching into a network of wires. We show that transport through the wires is dominated by a few spots where the electrostatic potential forms a valley surrounded by two hills. When the tip is placed above these spots with a negative voltage, the conductance decreases strongly and can even drop to zero. In addition, the conductance does not decrease smoothly when the tip approaches these spots, but shows several oscillations, which are clearly revealed by sensitive transconductance measurements. These oscillations are interpreted as a signature of localized states with Coulomb blockade in quantum dots that form in the 2DEG local potential valley. By lowering locally the electron density close to zero under the tip, we indeed expect the disordered potential landscape to form a series of potential barriers delimiting quantum dots with localized states, as depicted schematically in Fig.~\ref{fig1}. In the case of macroscopic 2D gates, it would give rise to the formation of isolated 2DEG islands and to a percolation-driven MIT. Our experiment therefore represents a local investigation of the disorder-induced MIT.

Similar conductance oscillations have been observed in SGM images of various systems, but were explained by the presence quantum dots only for InAs nanowires,~\cite{bleszynski-07-nl} carbon nanotubes,~\cite{woodside-02-sci} and graphene.~\cite{connolly-11-prb,pascher-12-apl,garcia-13-prb} So far, none of the SGM studies has found conductance oscillations due to the charging of quantum dots within a 2DEG but rather due to charging of traps or impurities in the heterostructure surrounding the conducting channel.~\cite{crook-02-prb,aoki-06-jpconf,pioda-07-prb,gildemeister-07-jap} Here, we show that the features observed in our experiment are not consistent with charging events in traps, but should instead be explained by the formation of quantum dots in the disordered potential landscape. We substantiate our finding by approximate electrostatic calculations of the disorder potential within the wire and the induced tip potential revealing nearly quantitative agreement with the experimental data.

The paper is organized as follows. Section~\ref{sec:exp1} gives technical information about the experiment. Sections~\ref{sec:exp2} and \ref{sec:exp3} present the SGM images and their analysis. Section~\ref{sec:sim1} presents simulations of the disordered potential landscape induced by ionized dopants. Section~\ref{sec:sim2} demonstrates that the SGM tip can reveal the presence of quantum dots and supports our analysis of the experimental data. Supplementary information, measurements, and analysis are given in Appendix sections.

\section{Experiment}\label{sec:exp}

\subsection{Sample and setup}\label{sec:exp1}

The sample is based on a pseudomorphic In$_{0.75}$Ga$_{0.25}$As/InAlAs heterostructure grown by molecular beam epitaxy on a semi-insulating InP substrate~\cite{wallart-05-jap} with the following layer sequence : 100\,nm lattice-matched InAlAs buffer layer, 50\,nm AlAsSb barrier, 400\,nm InAlAs layer, 15\,nm In$_{0.75}$Ga$_{0.25}$As channel, 20\,nm InAlAs spacer, $\delta$-doping Si plane ($2.25\times10^{12}$\,cm$^{-2}$), 15\,nm InAlAs barrier, 7\,nm doped InGaAs cap layer. The 2DEG is formed 42\,nm below the sample surface with a carrier density $n=3.5\times10^{11}$\,cm$^{-2}$ and a mobility $\mu=10^5$\,cm$^{2}$V$^{-1}$s$^{-1}$ as measured by magneto-transport at 4.2\,K in a Hall bar patterned on the same sample. The investigated nanostructure is a $1.0\times1.9\,\mu$m$^2$ network made of three 180\,nm wide parallel wires, linked together by two 210\,nm wide wires, connected to the source and drain reservoirs by 370\,nm wide openings (see Fig.~\ref{fig2}(a)). This complex sample geometry will be simply considered here as a set of three independent wires measured in parallel. The pattern is written by electron beam lithography and transferred into a mesa by wet etching of 65\,nm deep trenches.

SGM measurements are performed in a homemade atomic force microscope~\cite{martins-07-prl} (AFM) cooled at 4.2\,K by exchange gas in a liquid helium cryostat. A commercial silicon tip coated with a PtIr conducting layer is glued at the extremity of a tuning fork, which is used as a force sensor in the AFM imaging mode. Experiments start by recording a topographic image at 4.2\,K to locate the device. For SGM measurements, the tip is lifted by 100\,nm (for all the data presented here) and scanned in a plane at a constant height above the sample. Usually, a negative voltage relative to the 2DEG is applied to the tip, and the device conductance and/or transconductance are recorded as a function of the tip position during scanning. As a result of the capacitive coupling between the tip and the 2DEG, the electron density under the tip is reduced and the electrostatic potential is raised towards the Fermi level: the tip acts as a local movable gate.

The device conductance is measured with a lock-in using a small AC source-drain excitation at 68\,Hz, while a DC voltage is applied to the tip. The transconductance is measured with a small DC source-drain bias, while a 40\,mV AC excitation at 939\,Hz is applied to the tip in addition to the main DC voltage. The two signals can be recorded simultaneously with a dual reference lock-in amplifier. The unperturbed device resistance being around 10\,k$\Omega$, voltage or current bias can be used for the source-drain polarization, corresponding to the measurement of a conductance $G$ or a resistance $R$, respectively. Both configurations have been used depending on the highest resistance recorded in the SGM map, but all data are plotted here in terms of conductance and transconductance, using the conversion $G=1/R$ and ${\rm d}G/{\rm d}V_{\rm tip}=-(1/R^2)\,{\rm d}R/{\rm d}V_{\rm tip}$. Note that ${\rm d}G$ instead of ${\rm d}G/{\rm d}V_{\rm tip}$ will be plotted for the transconductance signal in order to keep all quantities ($G$ and ${\rm d}G$) in units of $2e^2/h$.

\subsection{Conductance images}\label{sec:exp2}

\begin{figure}[b]
\begin{center}
\includegraphics[width=8cm,clip,trim=0 0 0 0]{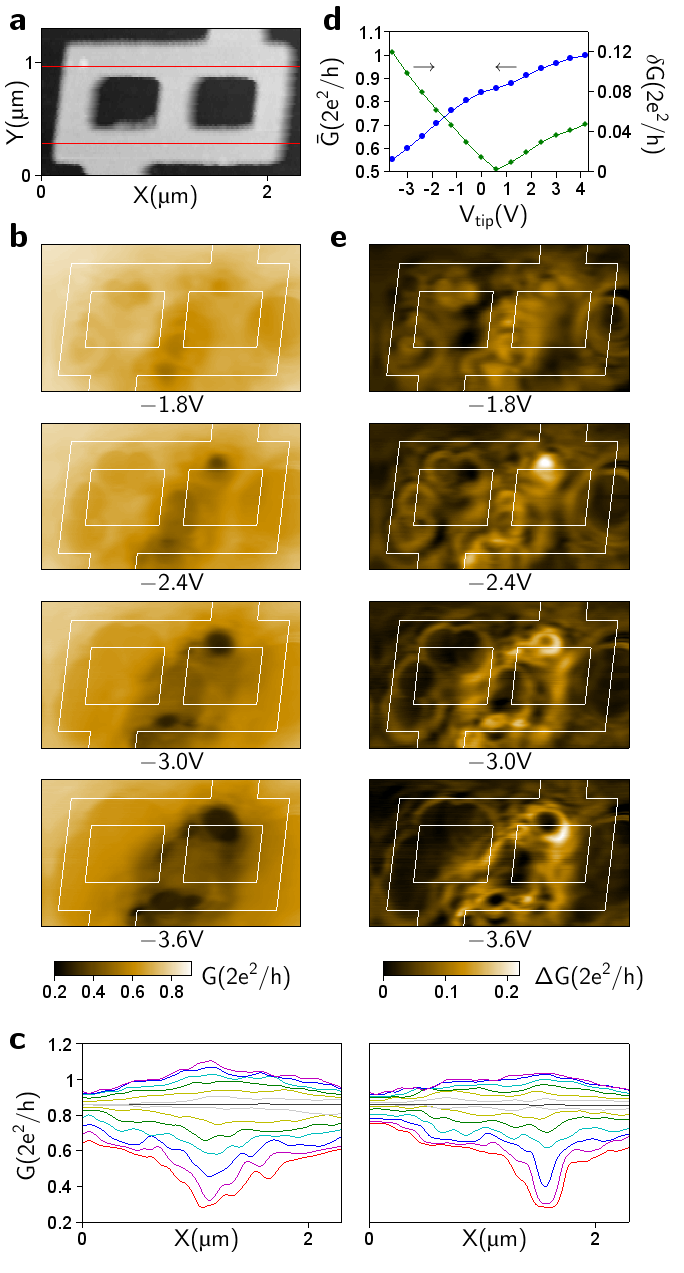}
\caption{(a) Topography at 4.2\,K recorded before the SGM measurements. (b) SGM images of the conductance $G$ measured with an AC current bias $I=10$\,nA. The DC tip voltage is indicated on each image. (c) SGM conductance profiles along the red lines drawn in (a) for tip voltages from $-$3.6\,V (bottom curve) to +4.2\,V (top curve) with 0.6\,V steps. Left and right graphs correspond to bottom and top red lines, respectively. (d) Average conductance $\bar{G}$ and standard deviation $\delta G$ calculated from the SGM images in (b) versus tip voltage. (e) Difference $\Delta G$ between two consecutive SGM images as explained in the text.} \label{fig2}
\end{center}
\end{figure}

The conductance images shown in Fig.~\ref{fig2}(b) are obtained by scanning the tip above the entire device for decreasing negative tip voltages. They show a complex pattern of conductance drops covering the device area between the two openings. The device geometry can hardly be recognized because the tip-induced potential has a broad lateral extension and influences electron transport in the device even if the tip is not directly above the wires. The largest changes are observed along the central path, which probably carries the largest current, and in particular at its ends which are critical nodes for transmission. At some locations, the conductance drops by a factor of 4 at $V_{\rm tip}=-3.6$\,V and can even drop to zero at larger negative tip voltage as shown later. The narrow width of the arms and the low electron density make the device very sensitive to potential changes induced by the tip.

SGM profiles recorded along two selected lines are plotted in Fig.~\ref{fig2}(c) for different tip voltages. It is found that the profiles recorded at $V_{\rm tip}^{\rm flat}=+0.6$\,V (black curves) show no conductance change, i.e., the tip does not produce any potential perturbation. This particular value, the so-called flat band voltage, corresponds to the work function difference between the PtIr coating of the tip and the InGaAs cap layer of the heterostructure taking into account a surface Fermi level pinning at mid-gap (see Appendix A). Similar values were found for PtIr tips and GaAs surfaces in other SGM experiments~\cite{vancura-03-apl,pioda-04-prl} since InGaAs and GaAs have similar work functions.

The average conductance calculated over the full scanning area is shown in Fig.~\ref{fig2}(d). It varies roughly linearly with the tip voltage as in the case of a macroscopic field effect transistor, except for the lower slope observed at positive voltages that we attribute to a larger screening in case of charge accumulation. The standard deviation of the conductance maps, also shown in Fig.~\ref{fig2}(d), is found to drop very close to zero at the flat band voltage $V_{\rm tip}^{\rm flat}=+0.6$\,V, showing that the tip is free from charged dust particles that would have disturbed its local gate action.~\cite{gildemeister-07-prb} The linear increase of the standard deviation on both sides of $V_{\rm tip}^{\rm flat}$ is consistent with an in-average linear gate effect of the tip (linear response), since the conductance does not drop to zero in this tip voltage range.~\cite{martins-07-prl} 

Careful examination of the conductance images in Fig.~\ref{fig2}(b) reveals that they contain several spots, growing in size and amplitude for decreasing tip voltages. The edge of these spots can be made more visible in Fig.~\ref{fig2}(e) by plotting the difference $\Delta G(V_{\rm tip})=G(V_{\rm tip}+0.6)-G(V_{\rm tip})$ between conductance maps recorded at two consecutive voltages separated by 0.6\,V. Many overlapping circles appear in these images, four in the left branch, two in the right one, and even more in the central one, which are difficult to distinguish. Their diameter increases for more negative tip voltages, but their center remains at a fixed position, always located inside the wires, never in the etched regions. These images show that the device is very sensitive to a local potential change at these particular locations.

Similar isolated features were observed previously in SGM images by other groups and were interpreted as the presence of charged traps in the semiconductor heterostructure, possibly in the doping plane, but not in the 2DEG itself.~\cite{crook-01-jp,crook-02-prb,aoki-06-jpconf,dacunha-06-apl,pioda-07-prb,gildemeister-07-jap} In this interpretation, the trapped charges create potential perturbations in the potential landscape of the 2DEG, and changing the number of these charges modifies the device conductance. In this case, approaching the tip with a negative voltage removes electrons from the traps and restores a larger device conductance.

In our case, the exact opposite behavior is observed, since approaching the tip with a negative voltage strongly decreases the conductance. The phenomenon observed in our experiment is therefore not related to traps in the heterostructure, and we propose instead that the tip affects directly the 2DEG potential within the following mechanism. When the tip scans above a high hill of the potential landscape, the gate effect of the tip is stronger, and it produces a spot of low conductance in the SGM image. Low density regions have indeed a weak screening capability and can be easily depleted by the repulsive potential of the tip. In addition, our particular device geometry composed of narrow wires makes the conductance very sensitive to a local depletion of the 2DEG. According to this proposal, which will be sustained later in the paper, SGM images reveal the spatial inhomogeneity of the 2DEG and show that it is characterized by a discrete distribution of small regions where the electron density is much lower than the average value.

Some of the features in Fig.~\ref{fig2}(e) consist of two concentric circles that may arise either from two very close spots, or from a single spot with two successive changes. To distinguish these two possibilities, we need higher resolution images. For this purpose, we now present transconductance measurements, which are more sensitive than a simple difference between two conductance maps.

\subsection{Transconductance images}\label{sec:exp3}

\begin{figure}[b]
\begin{center}
\includegraphics[width=\columnwidth,clip,trim=0 0 0 0]{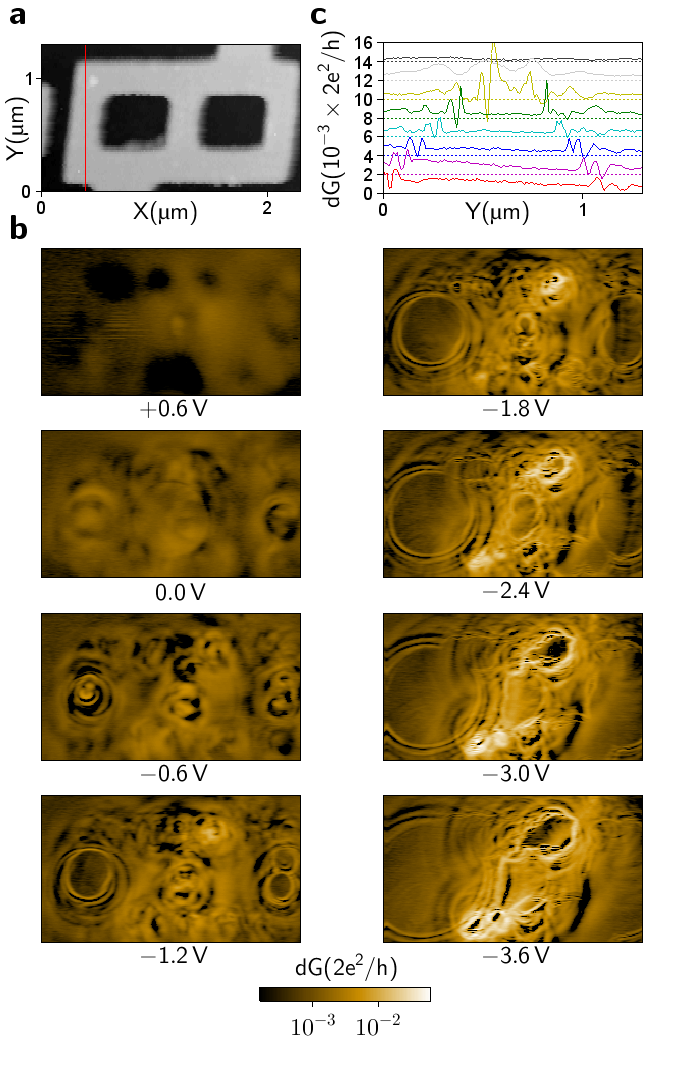}
\caption{(a) Topography at 4.2\,K recorded before the SGM measurements. (b) SGM images of the transconductance ${\rm d}G/{\rm d}V_{\rm tip}$ measured with a DC current bias $I=10$\,nA and an AC tip voltage modulation ${\rm d}V_{\rm tip}=40$\,mV. The DC tip voltage is indicated below each image. A logarithmic color scale is used and only positive values of the transconductance are plotted. (c) SGM profiles extracted along the red line drawn in (a). The successive profiles recorded from $-$3.6\,V to +0.6\,V are shifted upwards by $2\times 10^{-3}\times 2e^2/h$ for the sake of clarity (the dotted lines indicate the zeros).} \label{fig3}
\end{center}
\end{figure}

The transconductance signal ${\rm d}G/{\rm d}V_{\rm tip}$ is measured with a small additional AC voltage on the tip. A series of transconductance images is shown in Fig.~\ref{fig3}(b) for different tip voltages from $V_{\rm tip}^{\rm flat}$ down to $-$3.6\,V. Note that these images have been recorded during the same cool down as in Fig.~\ref{fig2}, but a small electrostatic discharge may have occurred during the change of the measurement configuration resulting in a slightly different potential landscape.

At $-$0.6\,V tip voltage, the SGM image shows several spots and circles which correspond to those visible in Fig.~\ref{fig2}(e). As clearly seen in the image at $-$1.2\,V, all these features are located along the device wires whose topography is shown in Fig.~\ref{fig3}(a). For more negative tip voltages, the spots evolve into narrow circles with increasing diameters. In the central wire, the presence of many overlapping circles makes the pattern rather complex to analyze. In the lateral wires however, only a limited number of spots dominate the conductance (one spot in the left wire, two spots in the right wire). Several concentric circles are visible around each spot, with at least two circles for each, and up to four circles for the left spot.

These circles look very much like the Coulomb blockade oscillations observed previously by SGM in different kinds of quantum dots made by lithography,~\cite{pioda-04-prl,kicin-05-njp,fallahi-05-nl} or present accidentally in nanowires,~\cite{bleszynski-07-nl} carbon nanotubes,~\cite{woodside-02-sci} and graphene.~\cite{connolly-11-prb,pascher-12-apl,garcia-13-prb} In our experiment, each spot showing concentric circles can therefore be interpreted by the presence of a quantum dot with Coulomb blockade. When the tip is approached towards the dot, the electrostatic potential is raised, which results in the discharging of electrons outside the dot, one-by-one, with a conductance maximum each time a charge state crosses the Fermi level. Because of the large electron density in the unperturbed 2DEG, these dots do not pre-exist in absence of the tip, but appear when the electron density is lowered under the tip, such that the potential fluctuations of the 2DEG are brought around the Fermi level. Fig.~\ref{fig1} illustrates this effect by showing a localized state under the tip with discrete energy levels close to the Fermi level. According to this interpretation, each set of concentric circles observed in the SGM images reveals the presence of a quantum dot formed in the 2DEG potential fluctuations. Note that the dots are not created by the tip, as done in the past with a scanning tunneling microscope on a clean InAs surface using a positive tip voltage.~\cite{dombrowski-99-prb} Here, the dots result from a local lowering of the density, such as to induce locally an equivalent to the disorder-induced metal-to-insulator transition, well-known in macroscopic 2DEGs at low enough electron density.~\cite{dassarma-05-prl}

These SGM images are reproducible within the same cool-down in absence of external perturbation, but change if light is shined on the sample or if an electrostatic discharge occurs in the setup. An example of images obtained after a small electrostatic perturbation is given in Appendix B. This behavior gives information about the origin of the electrostatic disorder, which is not structural, but results from a particular distribution of charges, located either in the doping plane, or at the surface, and which are frozen in a given configuration at low temperature.

A striking feature in the transconductance images is the appearance of a disk with constant signal inside the innermost circle, which grows in size for more negative tip voltage without any new circle appearing inside. This phenomenon is also visible in the SGM profiles recorded along a single line and plotted in Fig.~\ref{fig3}(c) : the transconductance oscillations are progressively shifted away from the center and a region with flat signal develops in the middle. Simultaneous conductance and transconductance measurements on a single constriction (see Appendix C) have shown that the absence of feature inside the innermost circle corresponds to zero current in the wire. This effect can be understood by the quantum dot being completely emptied and/or the barriers becoming too high to give significant tunneling. Since no current can flow around the dot (the wire is too narrow), the conductance of the wire vanishes. The total conductance of the device is however not zero because some current still flows in the two other wires of the network, and circles from quantum dots in these other wires are therefore visible inside the depleted areas (see Fig.~\ref{fig3}(b) below $-$2.4\,V in the left wire).

\begin{figure}[b]
\begin{center}
\includegraphics[width=\columnwidth,clip,trim=0 0 0 0]{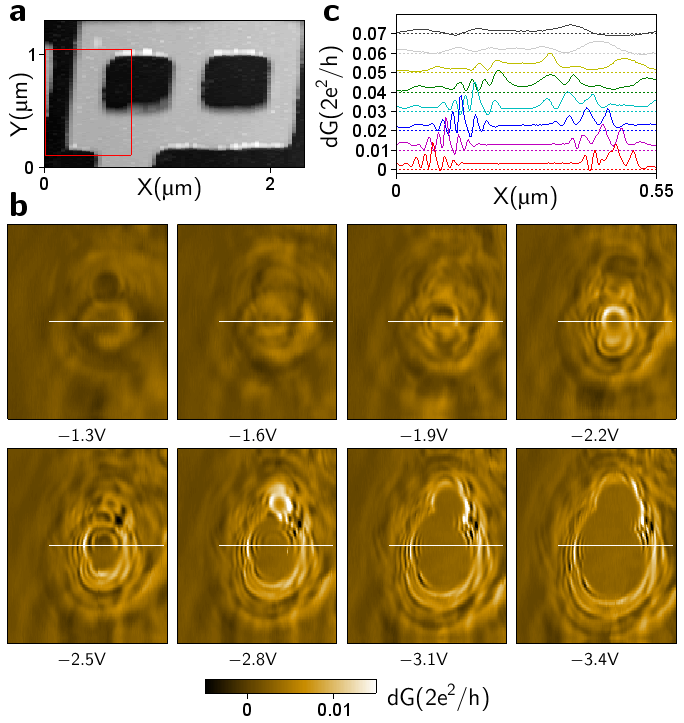}
\caption{(a) Topography of the device recorded before the SGM measurements. The red rectangle indicates the scanning area of the SGM images. (b) SGM images of the transconductance ${\rm d}G/{\rm d}V_{\rm tip}$ measured with a DC current bias $I=20$\,nA and an AC tip voltage modulation ${\rm d}V_{\rm tip}=40$\,mV. The DC tip voltage is indicated on each image. (c) SGM profiles extracted along the white line in (b). The successive profiles recorded from $-$3.4\,V to $-$1.3\,V are shifted upwards by $0.01\times 2e^2/h$.} \label{fig4}
\end{center}
\end{figure}

The concentric circles and their evolution with tip voltage are now analyzed in more details thanks to the high-resolution SGM images of the left wire plotted in Fig.~\ref{fig4}(b) for a slightly different disorder potential. Four dots in series with a similar response can be identified in this wire. Each dot shows a set of concentric circles, whose diameter increases for more negative tip voltages, and new circles emerge progressively from the center. Below a given tip voltage, a region of constant signal appears in the middle, corresponding to zero current in the wire (in parallel with the two other wires). This uniform region grows in size and merges progressively with the uniform regions of the nearby dots. Careful examination of these areas without signal shows that they are not exactly centered on the dots, and sometimes, cover partly the innermost circles. This indicates that a uniform region may not correspond to an empty dot with the last electron being removed, but rather to the appearance of a thick barrier around the dot that suppresses the current.

The SGM profiles in Fig.~\ref{fig4}(c) show the conductance oscillations for a single dot. About six oscillations are visible on both sides of the flat region where the wire is blocked. For a correct interpretation of the data, it is important to note that the transconductance signal corresponds to the derivative of the conductance curve with respect to gate voltage. Each Coulomb blockade conductance peak therefore appears as a transconductance oscillation with a negative peak immediately followed by a positive peak when the tip approaches the dot. Each oscillation is progressively shifted outwards the center when the tip voltage is decreased. The shift versus tip voltage is linear below $-$2.2\,V, which indicates an unscreened tip-induced potential. A detailed discussion of the potential induced by the tip in the wire in presence of screening effects is given in Appendix D. 

In our experiment, the successive Coulomb resonances are not separated by Coulomb-blocked regions, indicating that the charging energy is smaller than the temperature or the intrinsic resonance width, probably broadened by the poor confinement of the disorder potential. A tip-dot capacitance $C_{\rm tip,dot}=e/\Delta V_{\rm tip}=8\times 10^{-19}$\,F can be deduced from the tip voltage change $\Delta V_{\rm tip}=0.2$\,V required to shift the Coulomb oscillations by one period. Note that the transconductance is measured with a 40\,mV tip voltage modulation smaller than $\Delta V_{\rm tip}$ in order to fully resolve the Coulomb oscillations. The determination of the charging energy, however, requires the knowledge of the total dot capacitance, usually measured by source-drain bias spectroscopy of the dot. The presence of several dots in series between source and drain in this device prevents the investigation of an individual dot.

\section{Simulations}\label{sec:sim}

In this section, we develop a simple model to show that isolated dots hosting a few electrons can appear under the SGM tip due to the disordered potential landscape inside the narrow wire.

\subsection{Potential fluctuations}\label{sec:sim1}

Potential fluctuations in the 2DEG have several origins, including alloy disorder, fluctuations of the barrier thickness, random distribution of ionized dopants, inhomogeneous density of surface charges. For simplicity, we consider only the distribution of ionized dopants as source of potential fluctuations, since it is an intrinsic source that cannot be suppressed. Following Ref.~\cite{nixon-90-prb}, we calculate the potential induced by positively charged ions distributed randomly in a plane located at a distance $h=20$\,nm from the 2DEG, with a mean density $N_d=2\times10^{16}$\,m$^{-2}$. We use a boundary condition with a uniform potential on the surface located at a distance $p=40$\,nm from the 2DEG.~\cite{footnote} We assume the Fermi level to be pinned at mid-gap by the surface states of the InGaAs cap layer, such that the conduction band edge of InGaAs is at an energy $V_s\approx400$\,meV above the Fermi level. For simplicity, the dielectric constant is taken uniform over the heterostructure, using the value $\epsilon_r=12.7$ of the InAlAs barrier.

In the case of a uniform dopant distribution with a continuous density $N_d$, the positively charged dopants induce an attractive potential energy $V_d=-e^2N_d(p-h)/\epsilon_0\epsilon_r$ for electrons located below the doping plane with respect to the fixed surface potential. Therefore, electrons accumulate at the InGaAs/InAlAs interface and form a 2DEG with a uniform density $N_e$, which in turn induces a repulsive potential energy $V_e=e^2N_ep/\epsilon_0\epsilon_r$ in the 2DEG plane. The electron density $N_e=(m^*/\pi\hbar^2)(-V)\theta(-V)$ depends self-consistently on the total potential energy $V=V_s+V_d+V_e$ calculated with respect to the Fermi level ($m^*$ is the electron effective mass and $\theta$ is the Heaviside step function). The condition for a non-zero electron density in the 2DEG is having $V$ negative, such that the conduction band edge is below the Fermi level (which is set to zero as the energy reference). When the density is non-zero, the self-consistent potential energy writes:
\begin{eqnarray*}
 V = \frac{1}{1+\frac{m^*}{\pi\hbar^2}\frac{e^2\,p}{\epsilon_0\epsilon_r}}\;\left(V_s+V_d\right)
\end{eqnarray*}
where the coefficient before the parenthesis represents the screening by the 2DEG. For the chosen heterostructure parameters, this coefficient equals 0.09 and the attractive potential energy of the dopants is $V_d=-570$\,meV. In order to reproduce the measured electron density $N_e=3.5\times10^{15}$\,m$^{-2}$, we have to set the surface potential to $V_s=350$\,meV, which is close to the value expected from a Fermi level pinning at mid-gap on the surface.

\begin{figure}[b]
\begin{center}
\includegraphics[width=\columnwidth,clip,trim=0 0 0 0]{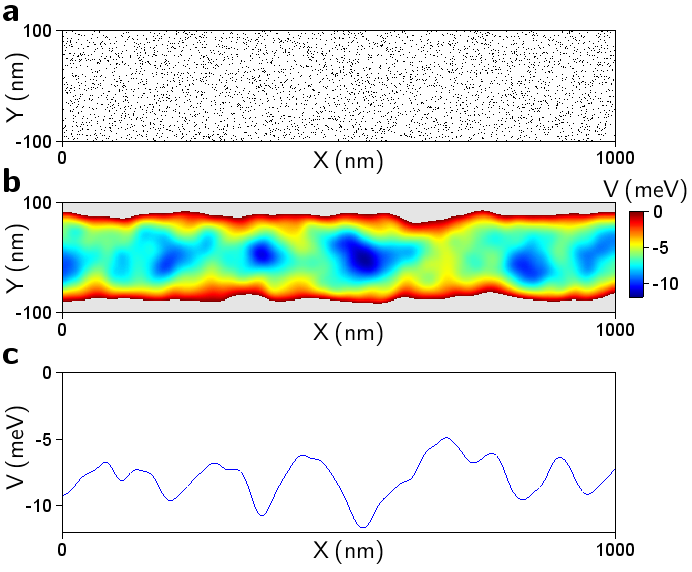}
\caption{(a) Random distribution of ionized dopants in the doping plane at finite distance above the 2DEG. The wire is 200\,nm wide and infinitely long. (b) Spatial fluctuations of the electron potential energy induced by the distribution in (a). The Fermi level is at $V=0$. The regions in gray are depleted. (c) Energy potential profile along the central line at $Y=0$. The Fermi energy is about 8\,meV with peak-to-peak fluctuations as large as 6\,meV.} \label{fig5}
\end{center}
\end{figure}

In the case of a random distribution of ionized dopants, the attractive potential energy $V_d$ is non-uniform and writes:
\begin{multline*}
 V_d(\vec{r}) = \frac{-e^2}{4\pi\epsilon_0\epsilon_r}\;\sum_i \left[\frac{1}{\left(|\vec{r}-\vec{r_i}|^2+h^2\right)^{1/2}}\right.\\ \left.-\;\frac{1}{\left(|\vec{r}-\vec{r_i}|^2+(2p-h)^2\right)^{1/2}}\right]
\end{multline*}
where the fixed surface potential is equivalent to the presence of an image charge with opposite sign. In this case, the electron density $N_e(\vec{r})$, the repulsive self-energy $V_e(\vec{r})$, and the total potential energy $V(\vec{r})$ are also non-uniform, and their exact determination would require self-consistent quantum calculations in the 2DEG plane. Here, we keep the calculation classical, and make the approximation of a local response using the same relations as for the uniform case. This is a first-order approximation to give an estimate of the potential fluctuations.

Fig.~\ref{fig5}(a) shows a random distribution of ionized dopants in a 200\,nm wide and infinitely long wire, while Fig.~\ref{fig5}(b) shows the resulting screened potential energy in the 2DEG. The finite width of the wire results in a larger attractive potential in the central region and 20\,nm wide depleted regions on each side (in gray). The Fermi energy in this wire geometry (about 8\,meV in the center) is significantly lower than the value for the infinite plane (20\,meV). The most striking property is the presence of strong potential fluctuations along the wire (Fig.~\ref{fig5}(c)) with peak-to-peak variations (about 6\,meV) of the same order as the Fermi energy (about 8\,meV). These fluctuations are proportional to the square root of the mean dopant density~\cite{davies-89-prb} and their typical length scale (about 50\,nm) is governed by the distance between the doping plane and the 2DEG.~\cite{nixon-90-prb} This length scale indeed corresponds to the extension of the potential induced by each dopant, which is much larger than the mean dopant spacing (7\,nm).

\subsection{Formation of quantum dots}\label{sec:sim2}

\begin{figure}[b]
\begin{center}
\includegraphics[width=\columnwidth,clip,trim=0 0 0 0]{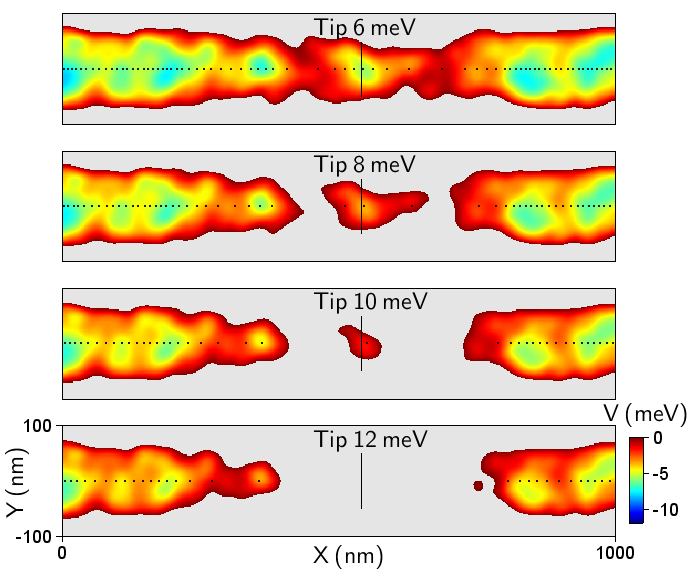}
\caption{(a-d) Potential energy landscape in the wire with the SGM tip at position $X_{\rm tip}=540$\,nm (indicated by the vertical bar) and $Y_{\rm tip}=0$\,nm. The dopant distribution is the same as in Fig.~\ref{fig5}. The spatial extension of the tip-induced potential is given in the text and its amplitude is increased from 6 to 12\,meV as indicated in each panel. The tip creates a low density region and a small quantum dot is formed for potentials between 8 and 10\,meV with a small number of electrons (indicated by black dots).} \label{fig6}
\end{center}
\end{figure}

In SGM experiments, for large enough negative tip voltage, the tip-induced potential brings locally the total electron potential above the Fermi level and builds a barrier that blocks electron transport along the narrow wire. In presence of potential fluctuations with a local minimum under the tip, a pocket of electrons can survive in this barrier, forming a small dot between two barriers as drawn schematically in Fig.~\ref{fig1}(c). When these confining barriers are rather symmetric, a resonant tunneling process through the dot can restore a high electron transmission for discrete energy levels. If the resistance of the barriers is larger than $h/e^2$, Coulomb blockade will also occur with charge quantization in the dot and a finite energy spacing between successive charge states. In this case, if the temperature is lower than the charging energy $e^2/C_{\rm dot}$, discrete conductance peaks should appear as a function of the gate voltage on the tip.

Fig.~\ref{fig6} shows an example of the formation of such a quantum dot when the tip is placed right above a local potential minimum. Panels (a) to (d) show the evolution of the potential landscape in the wire when the potential energy under the tip is raised from 6 to 12\,meV. The shape of the tip-induced potential is chosen of the form $Z/(X^2+Y^2+Z^2)^{1/2}$ with $Z=140$\,nm as in the experiment, corresponding to an unscreened potential (see Appendix D). By increasing locally the potential energy, an island of electrons forms, then shrinks, and finally disappears. To quantify the number of electrons in this island, the total charge is calculated by integration of the electron density, and each additional charge $e$ along the $X$ axis is marked by a black dot. For example, three electrons are present in the dot for the 8\,meV tip-induced potential. These simulations show that isolated islands with a few electrons can indeed form under the tip in presence of potential fluctuations along the wire. This result supports our interpretation of the experimental data in terms of Coulomb blockade in quantum dots formed in the 2DEG disorder potential.

\section{Conclusion}

SGM has been used to investigate locally the disorder-induced potential fluctuations in a 2DEG patterned into narrow wires. The SGM images reveal that a few discrete spots dominate the total resistance, corresponding to hills in the potential landscape. In addition, several concentric circles appear in the transconductance images, which are very similar to those observed previously for real quantum dots in the Coulomb blockade regime. These features indicate the presence of localized states in the 2DEG, confined between two hills of the disorder potential, when the tip lowers locally the electron density. Additional characterizations of these dots should be done in the future, in particular source-drain bias spectroscopy of a single dot in a short constriction, to measure their charging energy and level spacing.

\section*{Acknowledgement}

This work has been supported by the French Agence Nationale de la Recherche (MICATEC project), by the Grenoble Nanosciences Foundation (Scanning-Gate Nanoelectronics project), by FRS-FNRS projects J.0067.13 and U.N025.14, by FRFC grant no. 2.4503.12, and by the ARC project "stresstronics". B.H. and F.M. are research associate and postdoctoral fellow of the FRS-FNRS, respectively.

\section*{Appendix A : Work function}

The work function of a semiconductor with Fermi level pinning at mid-gap is given by $W=\chi_e+E_g/2$ where $\chi_e$ is the electron affinity and $E_g$ the band gap. According to this formula, the In$_{0.53}$Ga$_{0.47}$As cap layer has a work function $W_{\rm InGaAs}=4.9$\,eV (similar to the value $W_{\rm GaAs}=4.8$\,eV for GaAs). The AFM tip (PointProbePlus from NanoSensors) coated with a layer of Pt$_{0.95}$Ir$_{0.05}$ alloy has a work function $W_{\rm PtIr}=5.4$\,eV, as measured by Kelvin probe force microscopy~\cite{heim-04-nl,kim-05-jkps,spadafora-10-nl} (note that $W_{\rm Pt}=5.6$\,eV and $W_{\rm Ir}=5.3$\,eV). The tip voltage $V_{\rm tip}^{\rm flat}$ that compensates for the work function difference between the tip and the surface (also called flat band potential) is therefore equal to $+0.5$\,V for an InGaAs surface ($+0.6$\,V for a GaAs surface). This value is consistent with the value $+0.6$\,V extracted from Fig.~\ref{fig2}(d).

\section*{Appendix B : Different configuration}

\begin{figure}[b]
\begin{center}
\includegraphics[width=\columnwidth,clip,trim=0 0 0 0]{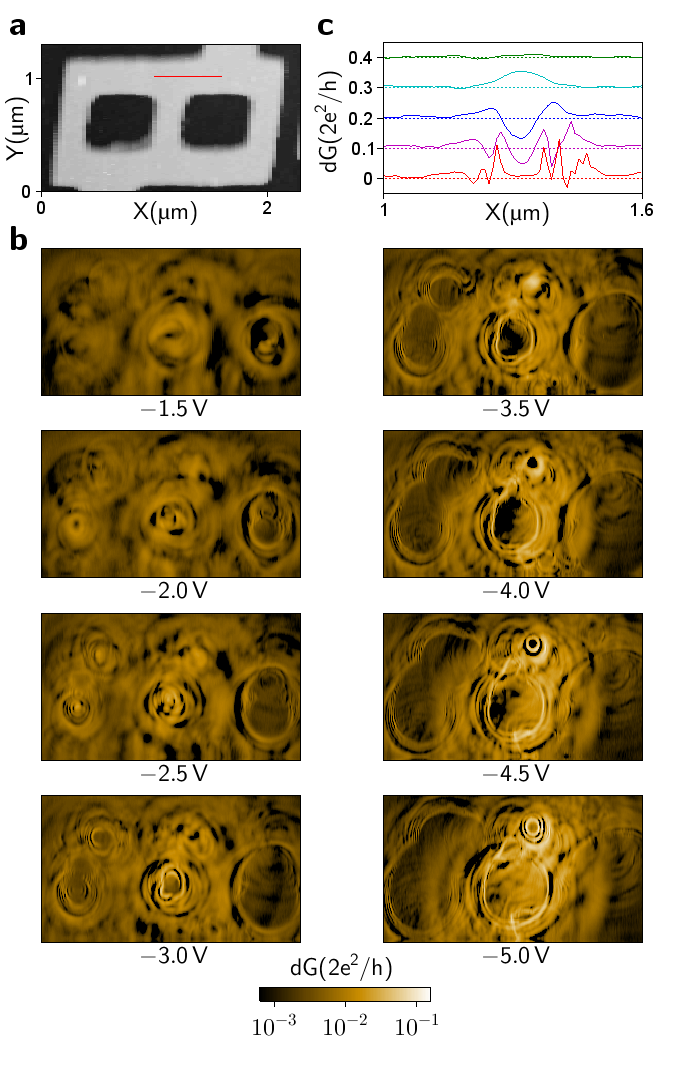}
\caption{Similar plots as in Fig.~\ref{fig3} for a slightly different disorder potential. (a) Topography recorded just before the SGM measurements. (b) SGM images of the transconductance ${\rm d}G/{\rm d}V_{\rm tip}$ measured with a DC current bias $I=20$\,nA and an AC tip voltage modulation ${\rm d}V_{\rm tip}=40$\,mV. The DC tip voltage is indicated on each image. (c) SGM profiles extracted along the red line drawn in (a). The successive profiles recorded from $-$5\,V to $-$3\,V are shifted upwards by $0.1\times 2e^2/h$ (the dotted lines indicate the zeros).} \label{fig7}
\end{center}
\end{figure}

Fig.~\ref{fig7} shows a set of SGM images recorded during the same cool-down as for Fig.~\ref{fig3} but after a refilling of the cryostat with liquid helium. Several dots can be recognized in the two sets of images, but some are new and others have disappeared. A small electrostatic discharge may have occurred in the cryostat during this operation, explaining a change of the charge distribution in the heterostructure, resulting in a slightly different potential landscape in the 2DEG. SGM profiles across a single dot in the upper wire are plotted in Fig.~\ref{fig7}(c). The conductance oscillations are rather large (amplitude up to $0.1\times 2e^2/h$) because this dot blocks the transport through two of the three wires of the device. When the tip voltage is lowered, the Coulomb peaks move away from the center and become sharper. This entails the faster potential change experienced by the dot when the tip is scanned with a larger negative voltage. At $-$5\,V tip voltage, the transconductance becomes flat and almost zero in the center because the local potential under the tip is so high that the current in the wire is completely blocked and the transconductance signal is suppressed.

\section*{Appendix C : Single constriction}

\begin{figure}[b]
\begin{center}
\includegraphics[width=\columnwidth,clip,trim=0 380 0 0]{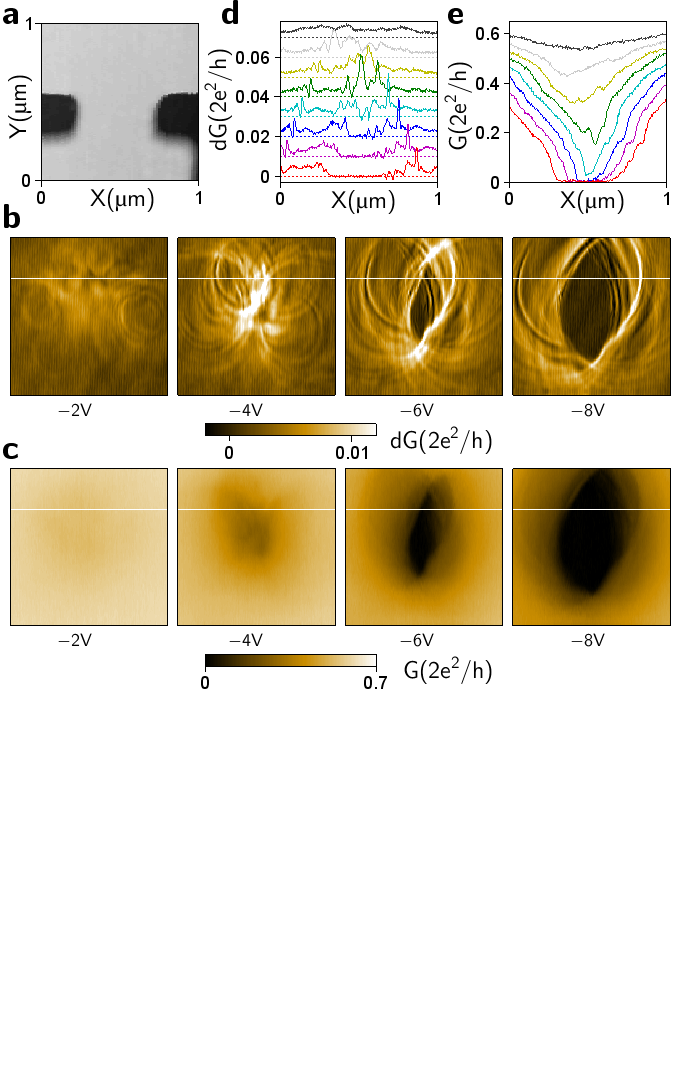}
\caption{(a) Topography of the upper constriction connecting the network to the top 2DEG reservoir. (b) SGM images of the transconductance ${\rm d}G/{\rm d}V_{\rm tip}$ measured with a DC voltage bias $V=1$\,mV and an AC tip voltage modulation ${\rm d}V_{\rm tip}=40$\,mV. The DC tip voltage is indicated on each image. (c) SGM images of the conductance $G$ measured with an AC voltage bias $V=100\,\mu$V simultaneously to the transconductance. (d,e) SGM transconductance and conductance profiles extracted along the white line in (b,c) for tip voltages from $-$2\,V to $-$9\,V (from top to bottom). In (d), the successive profiles are shifted by $0.01\times 2e^2/h$.} \label{fig8}
\end{center}
\end{figure}

A different configuration of the disorder potential was obtained by shining light on the sample at low temperature and waiting for charge noise relaxation. In this configuration, the upper device constriction shown in Fig.~\ref{fig8}(a) exhibits the strongest SGM response and dominates the device resistance. This situation corresponds to the presence of several negative charges in this region, which are frozen at the surface or in the doping plane, and raise locally the 2DEG potential. Simultaneous conductance and transconductance SGM measurements have been carried out in this region using a dual reference lock-in and plotted in Figs.~\ref{fig8}(b,c). Up to six dots can be identified in these images, arranged in parallel with respect to the current flow and controlling the amount of current flowing between the reservoir and the device. For tip voltages below $-$6\,V, a region with zero current and zero transconductance appears in the middle of the image, with a contour delimited by portions of different circles. This region corresponds to the overlap of the blocked regions created by the different dots. Individually, the dots cannot block the current because they are arranged in parallel rather than in series and several parallel paths are available for the current. Figs.~\ref{fig8}(d,e) show that weak conductance modulations correspond to sharp transconductance oscillations: when Coulomb blockade effects are weak, transconductance measurements strongly improve their detection in SGM images. On curves recorded with tip voltages lower than $-$7\,V , the region with a flat transconductance signal corresponds exactly to the region where the conductance is zero. This result shows that a flat transconductance signal usually indicates a vanishing current in the probed region.

\section*{Appendix D : Tip-induced potential}

\begin{figure}[b]
\begin{center}
\includegraphics[width=\columnwidth,clip,trim=0 0 0 0]{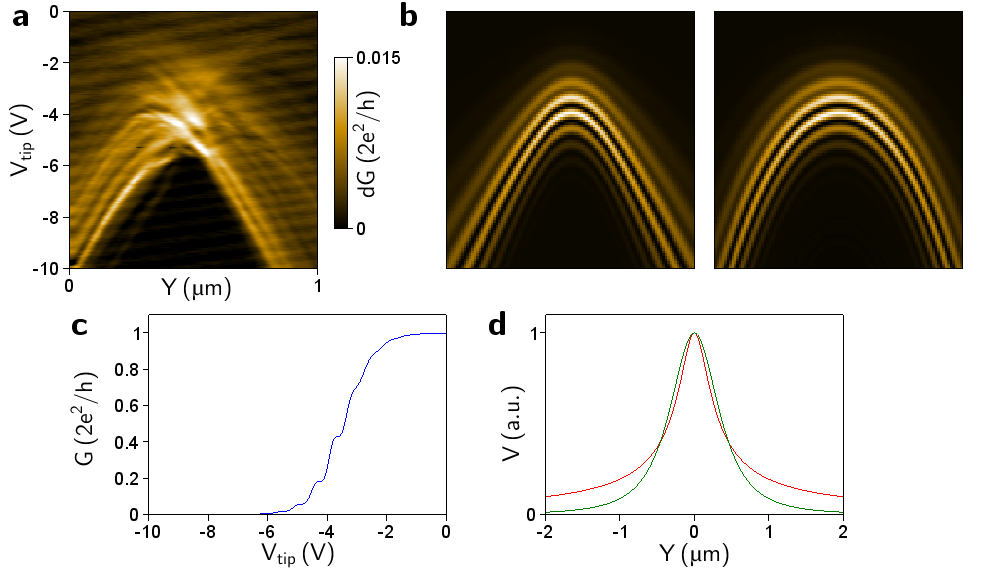}
\caption{(a) Single-line SGM scan along the $Y$ axis at $X=0.5\,\mu$m in Fig.~\ref{fig8}(a) for several tip voltages from 0 to $-$10\,V with 10\,mV steps. The transconductance ${\rm d}G/{\rm d}V_{\rm tip}$ is measured with a DC voltage bias $V=1$\,mV and an AC tip voltage modulation ${\rm d}V_{\rm tip}=40$\,mV (note that the weak tilted parallel lines covering the full plot are artifacts from a parasitic interference signal). (b) Simulation of the transconductance signal versus tip position and voltage (same axes as in (a)), using the modeled conductance curve shown in (c) and the tip-dot couplings shown in (d) without screening (left) and with screening (right). (c) Model for the quantum dot conductance versus tip voltage, when the tip is exactly above the dot. This curve simulates the gate effect around threshold and includes Coulomb blockade oscillations. (d) Model of the tip-dot coupling (or potential induced in the 2DEG plane) versus tip position, according to Eq.~\ref{eqn1} without screening (red line) and Eq.~\ref{eqn2} with screening (green line). The minimum tip-dot separation $\sqrt{X^2+Z^2}$ is 200\,nm and 500\,nm for the red and green lines, respectively.} \label{fig9}
\end{center}
\end{figure}

A direct measure of the energy change induced by the tip in the dots would require source-drain bias spectroscopy of an individual dot,~\cite{gildemeister-07-jap} but this study cannot be done here because of the multichannel character of the branched device. Alternatively, we investigate the potential induced by the tip in the 2DEG by measuring continuously the size of the concentric circles versus tip voltage while scanning a single line.~\cite{pioda-04-prl} Fig.~\ref{fig9}(a) shows the evolution of the transconductance signal along a vertical line in the middle of Fig.~\ref{fig8}(a) while sweeping the tip voltage. Following a given transconductance peak in this voltage-position diagram gives a trace $V_{\rm tip}^{\rm peak}(Y)$ that corresponds to an iso-potential line for the dot, i.e. a line where the potential induced in the dot is constant.~\cite{gildemeister-07-jap} From a theoretical point of view, the tip-induced potential along the $Y$ axis can be written: 
\begin{eqnarray*}
 V_{\rm induced}(Y) = \frac{C_{\rm tip,dot}(Y)}{C_{\rm dot}(Y)}\left(V_{\rm tip}-V_{\rm tip}^{\rm flat}\right)
\end{eqnarray*}
where $C_{\rm tip,dot}$ is the tip-dot capacitance, $C_{\rm dot}$ is the total dot capacitance, and $V_{\rm tip}^{\rm flat}$ is the flat band voltage. The quantity $C_{\rm tip,dot}/C_{\rm dot}$ represents the position-dependent lever-arm parameter between the tip voltage and the potential induced in the 2DEG. This parameter can be determined from an iso-potential line $V_{\rm tip}^{\rm peak}(Y)$, and then used to get the spatial dependence of the tip-induced potential at fixed tip voltage:
\begin{eqnarray*}
 V_{\rm induced}(Y) \propto \frac{V_{\rm tip}-V_{\rm tip}^{\rm flat}}{V_{\rm tip}^{\rm peak}(Y)-V_{\rm tip}^{\rm flat}}
\end{eqnarray*}
In Fig.~\ref{fig9}(a), it is not clear if the different traces correspond to different dots or to the successive charge states of the same dot, and the precise extraction of an iso-potential line is difficult. In the following, we adopt an alternative approach and compare the experimental figure with one resulting from the modeling of the potential induced by the tip in the dot with and without screening. 

For this purpose, we approximate the tip as a point charge $Q \propto V_{\rm tip}$ for which an analytic solution is possible. This charge is placed in vacuum above the surface of a semiconductor with dielectric constant $\epsilon_r$. The potential created inside a semiconductor, at coordinates $X,Y,Z$ relative to the charge, writes:
\begin{eqnarray}\label{eqn1}
 V(X,Y,Z) = \frac{Q}{4\pi\epsilon_0}\;\frac{2}{1+\epsilon_r}\;\frac{1}{\left(X^2+Y^2+Z^2\right)^{1/2}}
\end{eqnarray}
This expression is plotted as a red curve in Fig.~\ref{fig9}(d) for fixed values of $X$ and $Z$. If screening from the surrounding 2DEG can be neglected, this relation gives the potential variations of a dot inside the semiconductor, when a charge $Q$ is scanned in an horizontal plane above the surface with coordinates $X,Y,Z$ relative to the dot ($Z$ is the sum of the charge height above the surface and the dot depth below the surface).

To check if this unscreened $1/r$ dependence is consistent with the data in Fig.~\ref{fig9}(a), we simulate the transconductance signal in the presence of the tip. For this purpose, we model the dot conductance as shown in Fig.~\ref{fig9}(c), with a global drop due to the local depletion and weak Coulomb oscillations due to the disorder potential. This phenomenological model reproduces the typical behavior of the conductance curve versus gate voltage for a quantum dot in a disordered wire.~\cite{washburn-88-prb,staring-92-prb} The tip plays here the role of the gate. The left panel of Fig.~\ref{fig9}(b) shows the expected transconductance signal versus tip position and voltage. This plot reproduces qualitatively the experimental traces in Fig.~\ref{fig9}(a) if the minimum tip-dot distance is adjusted to $(X^2+Z^2)^{1/2}\sim 200$\,nm. Since the dot is in the 2DEG plane located at 42\,nm below the surface and the tip is at 100\,nm above the surface, i.e. $Z=142$\,nm, the horizontal distance between the scanning line and the dot is found to be $X\sim 140$\,nm. The linear asymptotic behavior of the experimental traces at large distance is well reproduced by this model without screening. Note that the successive charge states of the dot have different asymptotic slopes in the model, whereas the experimental traces are parallel to each other and may therefore correspond to different dots.

The above expression without screening predicts a very large tip-induced potential, which is not realistic. In reality, this potential is partially screened by the 2DEG, which is grounded to zero volt at the Ohmic contacts. Unfortunately, no analytical expression exists for the real case of a 2DEG embedded inside a semiconductor host and perturbed by a charge above the surface. In the following, we treat the closest situation as possible, which has an analytical solution, i.e. a 2DEG at the surface of the semiconductor. We therefore neglect the dielectric constant of the semiconductor barrier above the 2DEG and keep it only below the 2DEG. In the regime of linear response (no depleted region in the 2DEG) and in the Thomas-Fermi approximation (short Fermi wavelength), the potential in the 2DEG at a radial distance $r$ from a point charge $Q$ placed in vacuum at distance $Z$ above the 2DEG, can be calculated with the formula:~\cite{stern-67-prl,karsono-77-jpc,krcmar-02-prb}
\begin{eqnarray*}
 V(r) = \frac{Q}{4\pi\epsilon_0}\;\int_0^\infty{J_0(q\,r)\;e^{-\,q\,Z}\;\frac{2\,q}{(1+\epsilon_r)\,q+k_s}\;dq}
\end{eqnarray*}
where $k_s=m^*\,e^2/\pi\,\hbar^2\,\epsilon_0$ is the screening wave vector, $J_0$ is the zeroth-order Bessel function, and $\epsilon_r$ is the dielectric constant of the semiconductor located below the 2DEG. This formula is almost equivalent to the expression:
\begin{eqnarray*}
 V(r) = \frac{Q}{4\pi\epsilon_0}\;\frac{2}{1+\epsilon_r}\;\frac{1}{Z}\;\frac{I(a)}{1+a\,I(a)\left(\left(1+r^2/Z^2\right)^{3/2}-1\right)}
\end{eqnarray*}
where the integral $I(a)=\int_0^\infty{\frac{x\,e^{-x}}{x+a}\,dx}$ is a function of the dimensionless parameter $a=k_s\,Z/(1+\epsilon_r)$. Using $m^*=0.04\,m_e$ for InGaAs gives $k_s=3$\,nm$^{-1}$, then $\epsilon_r=14$ and $Z=142$\,nm give $a=28$. In this situation, $I(a)$ can be approximated by $1/a$ and the potential becomes:
\begin{eqnarray}\label{eqn2}
 V(r) = \frac{Q}{4\pi\epsilon_0}\;\frac{2}{k_s\,Z^2}\;\frac{1}{\left(1+r^2/Z^2\right)^{3/2}}
\end{eqnarray}
This expression is plotted as a green curve in Fig.~\ref{fig9}(d) and is independent of the semiconductor dielectric constant $\epsilon_r$ because of the large screening by the 2DEG located at the surface. This expression predicts a faster $1/r^3$ potential decay at large distance than Eq.~\ref{eqn1} without screening. The expected transconductance traces for this screened potential are shown in the right panel of Fig.~\ref{fig9}(b). Their nonlinear asymptotic behavior at large distance differ from the linear behavior of the experimental traces in Fig.~\ref{fig9}(a), which are better reproduced by the model without screening. This result might be explained by the very low electron density close to the depletion threshold where these traces have been measured.

\begin{figure}[b]
\begin{center}
\includegraphics[width=\columnwidth,clip,trim=0 0 0 0]{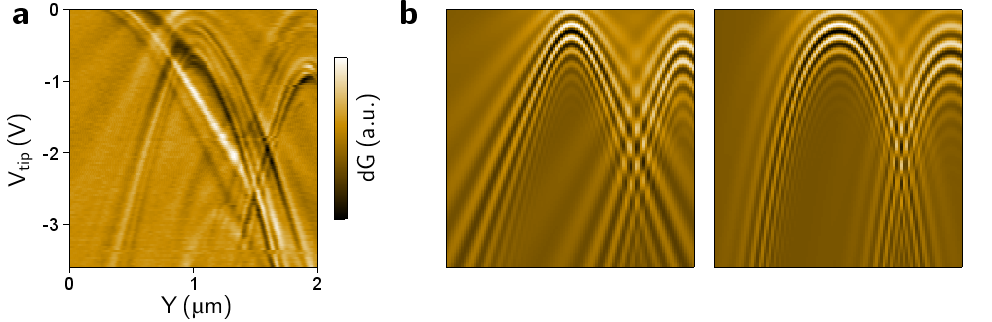}
\caption{(a) Single-line SGM scan along the $X$ axis in the middle of the device ($Y=0.65\,\mu$m in Fig.~\ref{fig2}(a)) for several tip voltages from $-$3.6 to 0\,V with 30\,mV steps. These data are recorded for a large device resistance (40\,k$\Omega$) after an electrostatic discharge. The transconductance ${\rm d}G/{\rm d}V_{\rm tip}$ is measured with an AC tip voltage modulation ${\rm d}V_{\rm tip}=40$\,mV. (b) Simulation of the transconductance signal versus tip position and voltage (same axes as in (a)), without screening (left) and with screening (right), like in Fig.~\ref{fig9}(b). Two conductance curves slightly different from that in Fig.~\ref{fig9}(c) are used to model two different dots contributing in parallel to the total conductance. The tip-induced potentials with and without screening are the same as in Fig.~\ref{fig9}(d).} \label{fig10}
\end{center}
\end{figure}

According to Eq.~\ref{eqn2}, screening by the 2DEG gives a reduction of the tip-induced potential by a factor $k_s\,Z \sim 400$, which gives a more realistic estimate of the potential in the 2DEG, as explained in the following. The charge $Q$ that dresses the SGM tip can be estimated using the sphere-plane capacitance model. The conical part of the tip above the apex also contributes to the tip-induced potential, but is not considered here. The tip is modeled by a metallic sphere of radius $R_{\rm tip}$ biased at a voltage $V_{\rm tip}$ relative to the grounded 2DEG at a gap distance $Z$. Its capacitance can be written~\cite{oyama-99-jap} $C=4\pi\epsilon_0\,R_{\rm tip}\,F(R_{\rm tip}/Z)$, where the function $F(x)\simeq(1+x)/(1+x/2)$ for $x<1$, $F(0)=1$, $F(1)=1.3$, $F(10)=2.1$. Since $R_{\rm tip}/Z<1$ for a sharp tip with small curvature radius, we can reasonably assume $F\simeq 1$. In this case, the charge is given by $Q/4\pi\epsilon_0\approx R_{\rm tip}\,V_{\rm tip}$. In this model, the screened potential in the 2DEG under the tip writes:
\begin{eqnarray*}
 V(0) = \frac{2\,R_{\rm tip}}{k_s\,Z^2}\;V_{\rm tip}
\end{eqnarray*}
This potential is of the order of 3\,mV for a 3\,V tip voltage and a 30\,nm tip radius (tip with metallic coating). Since the depletion of the 2DEG is obtained experimentally for a tip voltage of a few volts in the regions where dots are observed, we can estimate the Fermi energy to be about 3\,meV in these regions. This small energy is consistent with our simulation of the disorder potential in the wire (see Fig.~\ref{fig5}(c)) where a Fermi energy as small as 5\,meV is obtained on the highest potentiall hill. The existence of such high potential hills explains the strong response in the SGM images and the formation of quantum dots. Note that this model assumes an infinite 2DEG, whereas the device is etched into wires, which reduces the amount of screening as compared to the model.

For some disorder configuration, the different dots can be sufficiently far from each other to make the analysis of the dot characteristics easier. Fig.~\ref{fig10}(a) corresponds to such a case, with only three dots, one in each of the three parallel wires (the scanning line is in the $X$ direction along the symmetry axis of the device). This plot was recorded in very different conditions than for previous data, with a very large device resistance. About five traces are visible for the central dot and for the right dot. Two traces close to zero tip voltage come from a third dot in the left wire (this dot is rapidly depleted for negative tip voltages). In Fig.~\ref{fig10}(b), we compare these experimental traces with theoretical ones calculated using the above model. The figure shows the iso-potential traces for successive charge states of two different dots contributing in parallel to the conductance. The left panel corresponds to the unscreened tip-induced potential with $1/r$ decay (Eq.~\ref{eqn1}) and the right panel to the screened potential with $1/r^3$ decay (Eq.~\ref{eqn2}). The traces in the left panel reproduce better the shape of the experimental ones at large distance, as if the potential would not be screened. However, with Eq.~\ref{eqn1}, the tip voltage has to be artificially reduced by a factor 100 to give a realistic potential change in the 2DEG on the order of the Fermi energy, whereas Eq.~\ref{eqn2} gives reasonable values.

This analysis shows that screening effects are rather difficult to understand quantitatively in highly nonuniform systems like nanoscale devices, and would require 3D self-consistent calculations~\cite{szafran-11-prb} to be correctly taken into account. In addition, part of the discrepancy may result from the point charge model used for the tip, and numerical calculations would be necessary to treat correctly the actual shape of the tip.




\begin{thebibliography}{99}
\bibitem{footnote} This ``pinned surface'' boundary condition provides simple expressions for the potentials, but corresponds in principle to a room temperature situation where the surface states are conducting.
\end{thebibliography}
\end{document}